\documentclass{aa}  

\usepackage{graphicx}
\usepackage{txfonts}
\usepackage{placeins}

\newcommand{\kms}{\ensuremath{\rm km\,s^{-1}}}
\newcommand{\ms}{\ensuremath{\rm m\,s^{-1}}}

\begin{document}

   \title{Exoplanet atmospheres at high resolution through a modest-size telescope}

   \subtitle{Fe II in MASCARA-2b and KELT-9b with FIES on the Nordic Optical Telescope}

   \author{Aaron Bello-Arufe
          \inst{1}
          \and
          Lars A. Buchhave
          \inst{1}
          \and
          João M. Mendonça
          \inst{1}
          \and
          Ren\'e Tronsgaard
          \inst{1}
          \and
          Kevin Heng
          \inst{2,3,4}
          \and
          H.~Jens~Hoeijmakers
          \inst{5}
          \and
          Andrew W. Mayo
          \inst{6,7}
          }

   \institute{National Space Institute, Technical University of Denmark, Elektrovej, DK-2800 Kgs. Lyngby, Denmark\\
              \email{aarb@space.dtu.dk}
        \and
        University of Bern, Center for Space and Habitability, Gesellschaftsstrasse 6, CH-3012, Bern, Switzerland
        \and 
        University of Warwick, Department of Physics, Astronomy \& Astrophysics Group, Coventry CV4 7AL, United Kingdom
        \and
        Ludwig Maximilian University, University Observatory Munich, Scheinerstr. 1, Munich D-81679, Germany
         \and
            Lund Observatory, Box 43, Sölvegatan 27, SE-22100 Lund, Sweden
        \and
            Department of Astronomy, University of California Berkeley, Berkeley, CA 94720-3411, USA
        \and
            Centre for Star and Planet Formation, Natural History Museum of Denmark \& Niels Bohr Institute, University of Copenhagen, \O ster Voldgade 5-7, DK-1350 Copenhagen K., Denmark
             }

   \date{Received date; accepted date}

 
  \abstract
  {Ground-based, high-resolution spectrographs are providing us with an unprecedented view of the dynamics and chemistry of the atmospheres of planets outside the Solar System. While there is a large number of stable and precise high-resolution spectrographs on modest-size telescopes, it is the spectrographs at observatories with apertures larger than 3.5 metres that dominate the atmospheric follow-up of exoplanets. In this work, we explore the potential of characterising exoplanetary atmospheres with FIES, a high-resolution spectrograph at the 2.56 metre Nordic Optical Telescope. We observed two transits of MASCARA-2\,b (also known as KELT-20\,b) and one transit of KELT-9\,b to search for atomic iron, a species that has been recently discovered in both neutral and ionised forms in the atmospheres of these ultra-hot Jupiters using large telescopes. Using a cross-correlation method, we detect a signal of Fe II at the $4.5\sigma$ and $4.0\sigma$ level in the transits of MASCARA-2\,b. We also detect Fe II in the transit of KELT-9\,b at the $8.5\sigma$ level. Although we do not find any significant Doppler shift in the signal of MASCARA-2\,b, we do measure a moderate blueshift (3--6~\kms) of the feature in KELT-9\,b, which might be a manifestation of high-velocity winds transporting Fe II from the planetary dayside to the nightside. Our work demonstrates the feasibility of investigating exoplanet atmospheres with FIES, potentially unlocking a wealth of additional atmosphere detections with this and other high-resolution spectrographs mounted on similar-size telescopes.}


   \maketitle
%
\section{Introduction}
High-resolution spectroscopy is revolutionising our understanding of exoplanets. Over the past decade, this technique has allowed us to measure the speed of winds flowing across the terminator of exoplanets and probe thermal inversion layers and escape processes \citep[e.g.][]{snellen2010,nugroho2017,yanhenning2018}, and it has revealed the orbital inclination and rotation of different exoplanets \citep[e.g.][]{brogi2012,snellen2014}. Additionally, high-resolution spectroscopy has unleashed the detection of a profusion of molecular and atomic species in the atmospheres of various exoplanets \citep[e.g.][]{giacobbe2021,merritt2021,belloarufe2022}, and it is providing us with precise abundance measurements to understand where in the protoplanetary disk these planets formed \citep[e.g.][]{line2021}.

There are currently several optical and infrared spectrographs capable of studying exoplanet atmospheres at high spectral resolution \citep[see Table 1 of][for a non-exhaustive list]{fisher2020}. Given the minuscule size of the signal originating from an exoplanet atmosphere, these instruments are generally located at large telescopes, with typical aperture sizes between 3.5 and 11.8 metres.

In an attempt to demonstrate that high-resolution spectrographs on 2-m class telescopes can be used to characterise the atmospheres of exoplanets, \citet{kabath2019} observed WASP-18\,b using data from FEROS on the 2.2-m telescope at La Silla. Although they did not detect any spectral features, they predicted that the atmospheres of more favorable targets could be probed by high-resolution instruments on 2-m class telescopes. \citet{flagg2019} reported the detection of CO in CI Tau\,b using high-resolution spectra from IGRINS on the 2.7-m Harlan J. Smith Telescope. However, their detection also combined data from the same spectrograph installed on the larger 4.3-m Lowell Discovery Telescope.

In this work, we seek to demonstrate the capability of spectrographs at 2-m class telescopes to characterise exoplanetary atmospheres at high resolution. In particular, we use the FIbre-fed Echelle Spectrograph \citep[FIES;][]{telting2014}, a cross-dispersed high-resolution spectrograph mounted on the 2.56 m Nordic Optical Telescope \citep[NOT;][]{djupvik2010}, at the Observatorio del Roque de los Muchachos in La Palma, Spain. FIES has a maximum resolving power of $R\sim67,000$, and a spectral coverage that ranges from 3760~\AA\ to 8820~\AA.

As a proof of concept, we target MASCARA-2\,b (also known as KELT-20\,b) and KELT-9\,b \citep{gaudi2017,lund2017,talens2018}, two transiting exoplanets that have been subject to observational scrutiny by other high-resolution spectrographs \citep[e.g.][]{hoeijmakers2018,yanhenning2018, borsa2019,pino2020,rainer2021,kasper2021,paiasnodkar2022,yan2022}. MASCARA-2\,b and KELT-9\,b belong to the class of exoplanets known as ultra-hot Jupiters: gas giants with dayside temperatures $\gtrsim 2200$~K, hot enough to thermally dissociate most molecular species \citep{arcangeli2018,bell2018,kitzmann2018,parmentier2018,lothringer2018}. Our goal is to use FIES to replicate the detection of atmospheric neutral and ionised iron (Fe I and Fe II) in the transmission spectra of these extremely hot planets claimed by recent works \citep{hoeijmakers2018, hoeijmakers2019, hoeijmakers2020expres,casasayasbarris2019,cauley2019,stangret2020,nugroho2020,rainer2021}. Fe I and Fe II present multiple deep lines in the spectral range covered by FIES, so they should be ideal species to benchmark the efficiency of FIES/NOT against other instruments.




\section{Observations and Data Reduction}\label{sec:obs}
Using the high-resolution fibre of FIES ($R$\,$\sim$\,$67,000$), we observed two transits of MASCARA-2\,b\footnote{A third transit of MASCARA-2\,b was scheduled for the night of 7 Oct 2018, but it was lost due to poor weather conditions.}. The first transit took place on the night of 14 Oct 2018, which we refer to hereafter as Night 1. We observed the second transit on the night of 9 Sept 2019. We refer to this night as Night 2. We set the time of each exposure to 200 seconds in both nights, obtaining 63 exposures in Night 1 (42 of which were during transit) and 58 in Night 2 (38 of these in transit). To calibrate our spectra, we used ThAr exposures, taken every fifth science exposure in the case of Night 1, and at the beginning and end of the observations in Night 2. The airmass of the target was in the range 1.00--2.79 during Night 1, and 1.02--1.60 during Night 2.


We only observed one transit of KELT-9\,b, on the night of 18 Sept 2020, also with the high-resolution fibre. We used exposure times of 600 seconds which, after $\sim 6$ hours of observations, yielded 29 exposures, 18 of these in transit. In this case, we took a ThAr calibration after every third science exposure. The airmass of KELT-9 during the observations started at 1.05, decreased to 1.02 and eventually reached 1.87.
\begin{figure*}
\centering
\includegraphics[width=\linewidth]{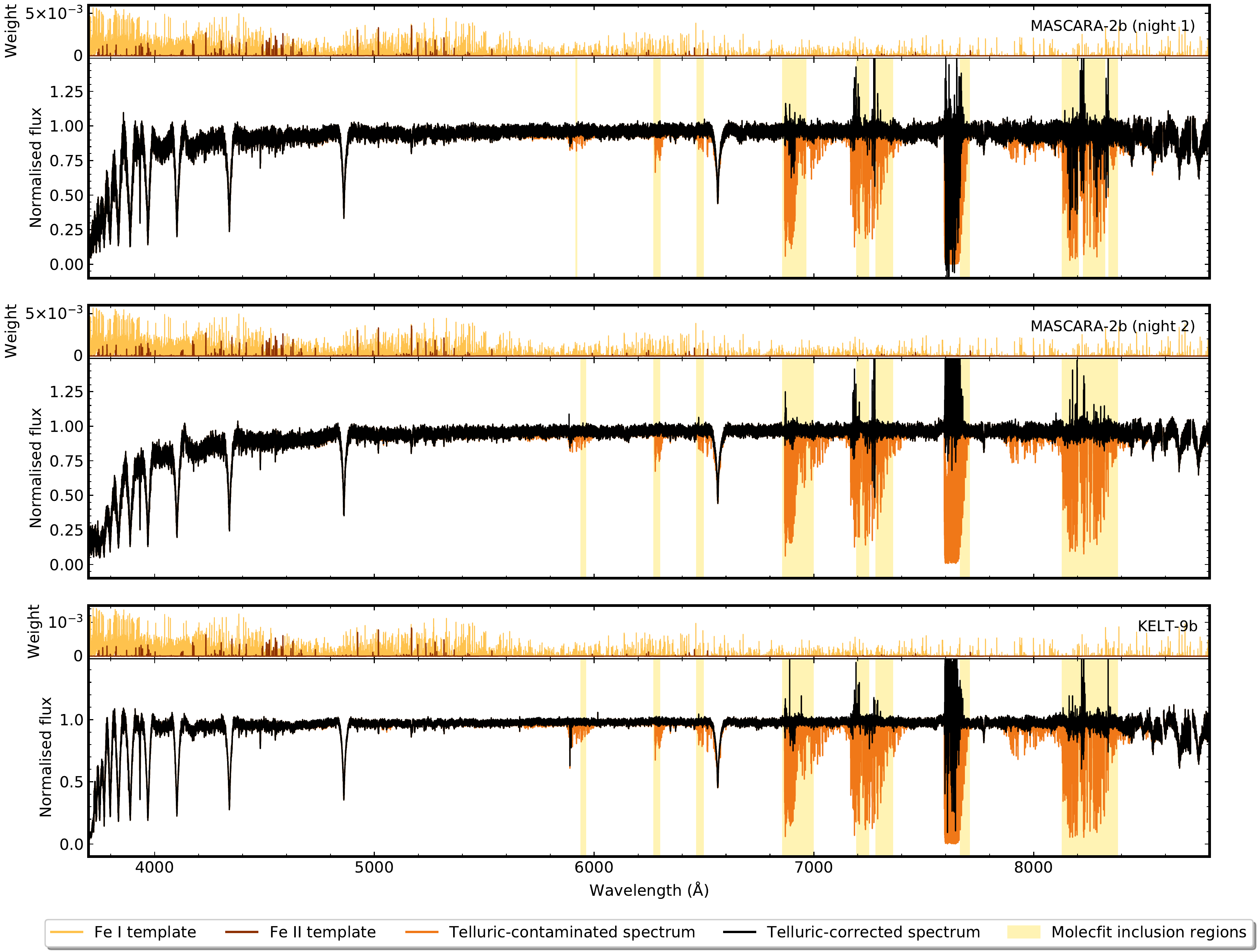}
  \caption{Three sample spectra (MASCARA-2\,b in the top and middle panels, KELT-9\,b in the bottom panel) before and after telluric correction with \texttt{molecfit}. Above each spectrum, we include the Fe I and Fe II model templates used in the cross-correlation analysis. While most Fe II spectral lines concentrate at short wavelenghts, telluric lines abound at long wavelengths.}
  \label{fig:telluric_correction}
\end{figure*}


We extracted the data as described by \citet{buchhave2010}, and we used \texttt{molecfit} to model and remove the telluric contamination in the spectra \citep{smette2015}. For the purposes of telluric correction, we deblazed, continuum normalised, and merged the echelle orders of each spectrum onto a uniform wavelength grid of 0.04~\AA\ steps. In the rest of our analysis, we used the individual echelle orders. As input for \texttt{molecfit}, we set the $\chi^2$ and parameter convergence criteria to $10^{-5}$. We allow \texttt{molecfit} to fit the spectral resolution of the data with a Gaussian function. We set the initial value for the FWHM of the Gaussian equal to 2.475 pixels, a choice that is based on the resolving power of the spectrograph ($R$\,$\sim$\,$67,000$) and the size of each pixel. We chose a value of 3000 for the KERNFAC parameter, which indicates the size of the Gaussian kernel in units of FWHM. We used polynomials of degree 3 and 2 for the continuum fit and wavelength calibration, respectively. Lastly, through quadratic interpolation, we shifted each telluric model produced by \texttt{molecfit} to the wavelength grid of the order-by-order spectra to correct the individual orders. Figure~\ref{fig:telluric_correction} illustrates the telluric correction of three sample merged spectra (two of MASCARA-2\,b and one of KELT-9\,b) and the spectral regions used in the fit. 

In general, we achieved a high-quality telluric correction, with the notable exception of some of the deepest water and oxygen lines found beyond 6800~\AA. This spectral region has a limited influence in the search for Fe II, as this species has very few lines at long wavelengths (see Fig.~\ref{fig:telluric_correction}). However, these poorly-corrected telluric lines can affect the search for Fe I, whose lines are present throughout the spectrum. Therefore, we masked out telluric lines in the following spectral regions during the analysis: 6866--7058~\AA, 7156--7374~\AA, 7593--7704~\AA, and 8101--8377~\AA. These wavelengths refer to air wavelengths in the observatory rest frame.

\section{Methods}
In our analysis, we used a cross-correlation approach to recover the signal from the exoplanet atmosphere, deeply buried in the noise of our data. This approach allows us to combine the signal from multiple spectral lines by cross correlating our data with a model template. In this section, we describe our cross-correlation analysis. We also describe how we produced the atmospheric models that acted as cross-correlation templates, and how we obtained an independent measurement of the systemic velocity using the stellar spectra. In our analysis, we treated each night independently, and only combined the results from the two nights of MASCARA-2\,b at the end.

\subsection{Cross-correlation analysis}\label{sec:cc_analysis}
The first step in our cross-correlation analysis was to place the telluric-corrected spectra in a reference frame where the exoplanet host star had a constant velocity. This step allowed us to remove the stellar contribution at a later stage. Originally, the spectra were in the rest frame of the observatory. Therefore, we Doppler shifted the wavelength solution of each exposure by the corrections due to the Earth's barycentric velocity and the radial velocity of the host star. The barycentric velocity correction accounts for the motion of the observatory around the Solar System barycentre. We calculated it using \texttt{barycorrpy} \citep{kanodiawright2018}, the Python implementation of the algorithm of \citet{wrighteastman2014}. The correction due to the host star radial velocity is given by
\begin{equation}
    v_{\star,\rm{corr}} = K_\star \sin{(2\pi\phi(t))},
\end{equation}
where $K_\star$ is the stellar radial velocity semi-amplitude and $\phi(t)$ is the exoplanet orbital phase. We did not correct for the stellar radial velocity of MASCARA-2 because its $K_\star$ value is only known to a 3$\sigma$ upper limit of $K_\star<311.3 ~ \ms$ due to its rapid rotation, $v\sin{i_\star} \simeq 120~\kms$ \citep{lund2017}. Given the level of rotational broadening of the stellar spectra and the width of a FIES pixel ($\sim 2~\kms$), not correcting for the stellar radial velocity of MASCARA-2 should not significantly impact our results \citep{casasayasbarris2018}. We then averaged the different wavelength solutions of each night to produce a common wavelength grid, and we shifted each spectrum to its corresponding grid using a quadratic interpolation.

The next steps were normalisation of the spectra and rejection of outlying flux values, for which we followed a strategy similar to that of \citet{hoeijmakers2020}. First, we normalised the flux in each order such that the mean of each order stayed constant throughout the night. As in \citet{hoeijmakers2020}, we preserved information about the average flux in each exposure by weighting each exposure accordingly at the end of our analysis (see Sect.~\ref{sec:kpvsys}). We then performed sigma-clipping on each order, using a sliding window of dimensions 40 pixels by $N_{\textup{exp}}$ columns (where $N_{\textup{exp}}$ is the number of exposures in each night of observation). We replaced values farther than 5 standard deviations away from the mean of the window by the linear interpolation of the adjacent pixels. We masked columns with more than 20\% of sigma-clipped values. We also masked those columns with excessive noise (i.e. with the following air wavelengths in the observatory rest frame: 3705--3708~\AA, 3729--3733~\AA, 3753--3757~\AA, and 3778--3780~\AA) or where the accuracy of the telluric correction was inadequate (i.e. the spectral regions listed in Sect.~\ref{sec:obs}). Finally, we normalised again the flux to ensure that the mean flux of each order remained constant throughout the night after the sigma-clipping and masking processes.

At this stage, the spectra were ready for the cross correlation. For each spectrum, we calculated its corresponding cross-correlation function as
\begin{equation}\label{eq:ccf}
    \textup{CCF}(v,t) = {\sum_{i=0}^{N_{\textup{pix}}}{f_i(t) T_i(v)}},
\end{equation}
where $f_i(t)$ is the flux at pixel $i$ and at exposure time $t$, and $T_i(v)$ is the value of the model template, Doppler shifted by a velocity $v$, at pixel $i$. The sum spans all $N_{\textup{pix}}$ pixels in each spectrum. We performed the cross-correlation at velocity shifts between $-500$ and $500~\kms$, in steps of $2~\kms$. The model template is normalised such that its sum over all pixels equals 1. As we did not continuum normalise the flux or divide it by the blaze function, Eq.~\ref{eq:ccf} naturally weights each pixel by its Gaussian noise.

We divided each in-transit CCF by a master CCF to remove the stellar signal and isolate the contribution from the transmission spectrum of the planet, and we then applied a broad Gaussian filter to remove any remaining low-frequency variations \citep{hoeijmakers2020}. In principle, we should have calculated the master CCF by averaging only the out-of-transit CCFs. However, in all three nights, we had a limited number of out-of-transit spectra, and these were mostly obtained at high airmass. Therefore, we calculated the master CCF of each night by averaging all exposures, in and out of transit. As both KELT-9\,b and MASCARA-2\,b are changing radial velocity rapidly, their signals spread over many pixels during the transit. Consequently, including the in-transit spectra as part of the master CCF would only marginally attenuate the planetary signals. And in our case, these attenuations are significantly outweighed by the reduction in noise and in stellar residuals gained from combining all exposures (e.g. the significance of the detection of Fe II in KELT-9\,b improves by 1.5$\sigma$). We also note that this method of including all CCFs to remove the stellar signal does not lead to uncorrected Rossiter-McLaughlin residuals in the final CCFs, since we are using a forward-modelling approach to account for this effect (which we describe in Sect.~\ref{sec:rm_clv_corr}).

\subsection{Atmospheric model templates}
To search for the presence of neutral and ionised atomic iron in KELT-9\,b and MASCARA-2\,b, we produced, for each planet, two model templates of each species. These model templates dictate the weight of each pixel in the cross correlation, according to Eq.~\ref{eq:ccf}. In this work, we produced these templates adopting the same methodology as \citet{belloarufe2022}, which we describe below.

We first calculated the gas opacities with \texttt{HELIOS-K} \citep{grimm2021} using the line list from \citet{kurucz2018}. We assumed Voigt line profiles, a line cutoff of 100~cm$^{-1}$, and a spectral resolution of 0.032~cm$^{-1}$, which is much higher than the instrumental resolution of FIES. We calculated the gas concentrations with \texttt{FastChem}, assuming chemical equilibrium and solar metallicity, and we incorporated H$^-$ bound-free and free-free absorption from \citet{john1988}.

To calculate the transmission spectra model templates, we divided the atmosphere into 200 annuli and calculated the wavelength-dependent transit radius following \citet{gaidos2017} and \citet{bower2019}. In our model, we adopted isothermal temperature profiles, setting the atmospheres of both planets to a uniform temperature of 4000~K \citep{hoeijmakers2018, stangret2020}. Prior to cross correlation, we removed the continuum from the templates, such that the template equaled zero everywhere except where spectral lines were present. We found the continuum by first using a sliding maximum filter to locate a set of anchor points, and then we applied a quadratic interpolation. As \citet{hoeijmakers2020}, in order to maintain a constant continuum, we set any point with amplitude smaller than $10^{-4}$ times the value of the deepest spectral feature to zero. Finally, we broadened each template to match the instrumental resolution of FIES. We present the model templates used in this analysis in Fig.~\ref{fig:telluric_correction}.

\subsection{Rossiter-McLaughlin and Centre-to-Limb Variation Effects}\label{sec:rm_clv_corr}
Cross correlating the spectra against a model template of a species that is present in the stellar atmosphere can lead to residuals in the CCFs due to the Rossiter-McLaughlin (RM) and centre-to-limb variation (CLV) effects. The RM effect occurs when a planet transits a rotating star: as the planet moves across the stellar disk, it blocks regions with different line-of-sight velocities, deforming the rotationally-broadened stellar lines. The CLV effect arises from differences in line profile and intensity between the centre and the limb of the host star. As our analysis targets Fe I and Fe II, two species present in both host stars, we must take these effects into consideration.

\begin{figure*}
\centering
\includegraphics[width=\linewidth]{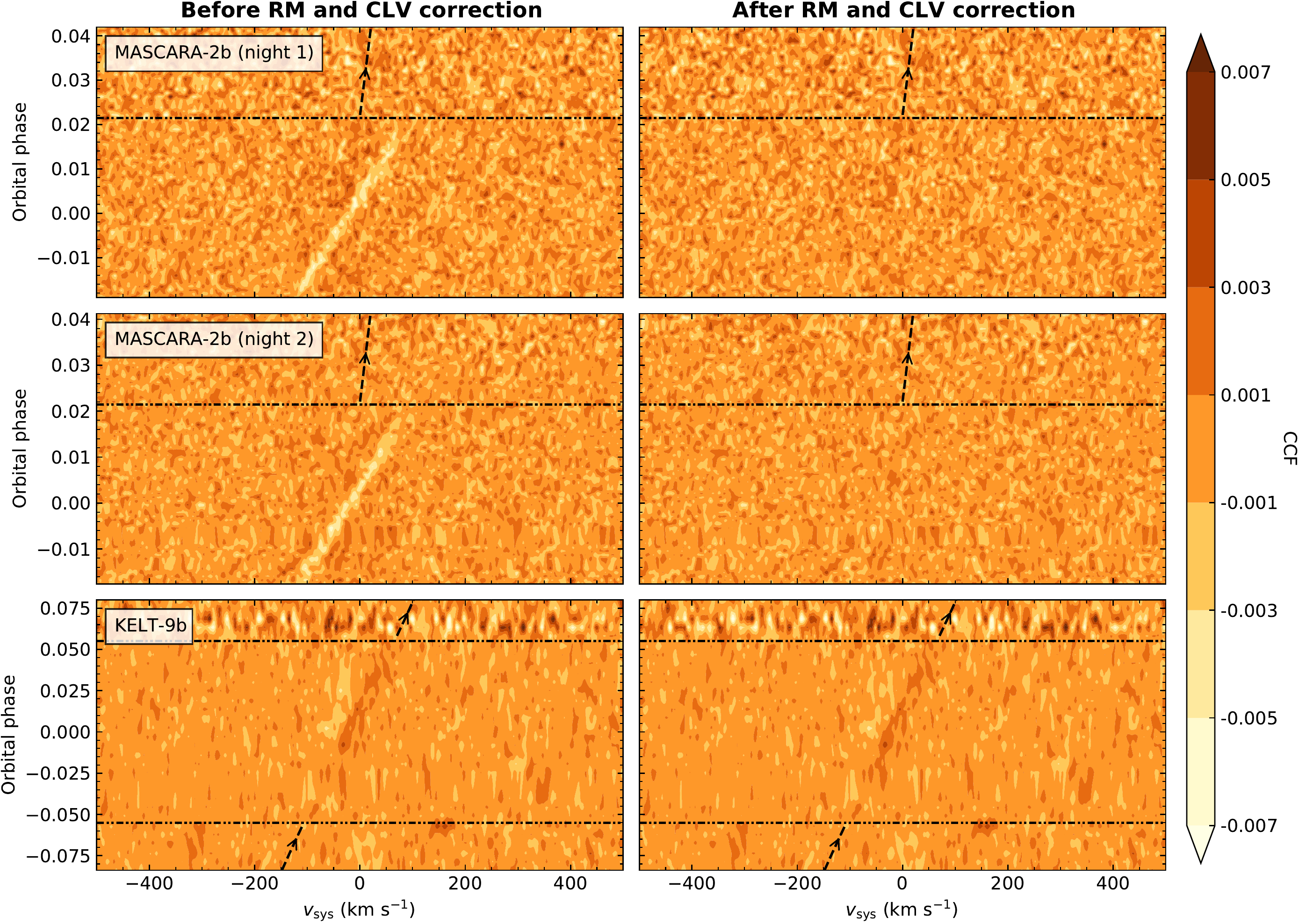}
  \caption{Stacked Fe II cross-correlation functions (CCFs), in the solar system's barycentric rest frame, of the three transits, before (left) and after (right) removing a model of the Rossiter-McLaughlin (RM) and center-to-limb variation (CLV) effects from the spectra. The dash-dotted lines mark the beginning and end of transit (we started observing the transits of MASCARA-2\,b shortly after they began). The slanted dashed lines with arrowheads trace the predicted radial velocity of the planet outside transit \citep{gaudi2017,lund2017}. While the RM and CLV contributions are evident as a bright yellow \textit{Doppler shadow} in the in-transit CCFs in the left pannels of the MASCARA-2\,b transits, they are not so obvious in the KELT-9\,b CCFs. This is because KELT-9\,b is on a polar orbit, and as we are using both in- and out-of-transit CCFs to calculate the master stellar CCF, the Doppler shadow largely vanishes. We note that the left pannels are just for the purpose of visualising the RM and CLV correction: we correct the spectra prior to cross correlation, so these contaminated CCFs are not part of our analysis. The atmospheric signal of Fe II from KELT-9\,b is already discernible as a brown streak tracing the planet's velocity in the in-transit CCFs.}
  \label{fig:ccfs_feii}
\end{figure*}

\begin{figure*}
\centering
\includegraphics[width=\linewidth]{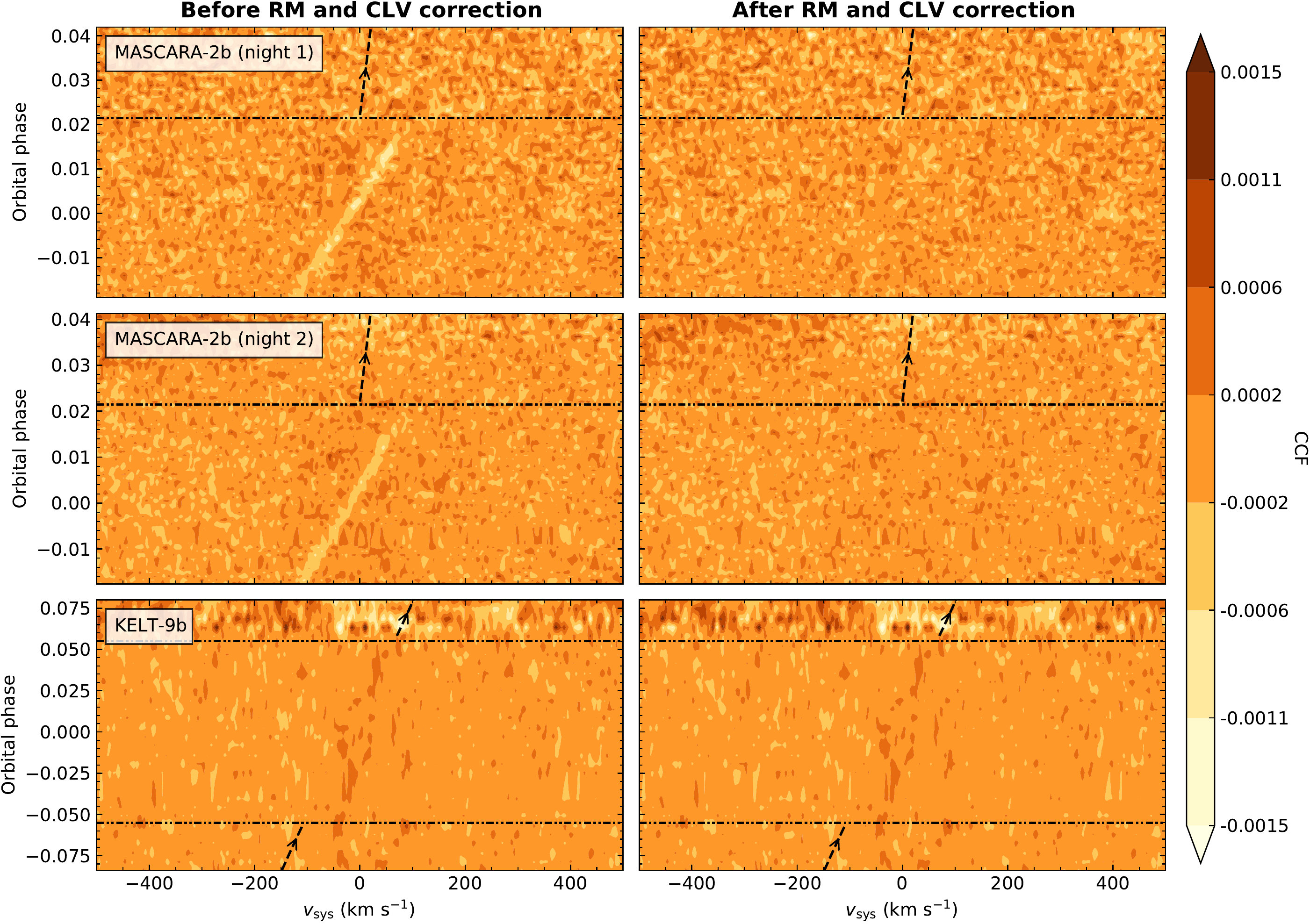}
  \caption{Same as Fig.~\ref{fig:ccfs_feii}, but for Fe I.}
  \label{fig:ccfs_fei}
\end{figure*}

We used forward modelling to remove the contribution from the RM and CLV effects \citep[e.g.][]{casasayasbarris2018,nugroho2020,belloarufe2022}. With \texttt{Spectroscopy Made Easy} version 522 \citep{piskunov2017}, we produced stellar spectra of each of the two exoplanet host stars at 21 different limb angles, between $\mu = 0.001$ and $\mu = 1$, using the the Kurucz \texttt{ATLAS9} models \citep{castelli2003} and the Vienna Atomic Line Database \citep{ryabchikova2015}. We divided the stellar disks into a grid of $0.01~R_\star\times0.01~R_\star$ cells, and we assigned to each cell a spectrum calculated from linear interpolation of the 21 precomputed spectra. Additionally, we Doppler shifted the spectrum in each cell according to the rotation of the star. At each phase, we integrated the spectra from the cells not blocked by the planet disk, and we divided the resulting integrated spectrum by the full-disk spectrum. Finally, we used a median filter to remove the continuum from each of these spectra, and we broadened them to match the instrumental resolution of FIES. This methodology allows us to obtain a model of the RM and CLV contributions at each orbital phase, which we then removed from the data. Figures~\ref{fig:ccfs_feii} and \ref{fig:ccfs_fei} show the CCFs that result from cross correlating the Fe II and Fe I templates with the three data sets before and after modelling the RM and CLV effects. These Figures evidence that our method achieves a correction down to the noise level.


\subsection{$K_{p}-v_{\rm sys}$ diagrams}\label{sec:kpvsys}
We produced $K_{p}-v_{\rm sys}$ diagrams to combine the signal from the in-transit CCFs of each night \citep{brogi2012}. Assuming a circular orbit, the radial velocity of the planet at time $t$ is given by
\begin{equation}\label{eq:vplanet}
    v_{p}(t) = K_p \sin{(2\pi\phi(t))} + v_{\textup{sys}},
\end{equation}
where $K_p$ is the planet radial velocity semi-amplitude, and $v_{\textup{sys}}$ is the systemic velocity. Therefore, for each pair of $K_p$ and $v_{\textup{sys}}$ velocities, we averaged the values of the in-transit CCFs at velocities $v_{p}(t)$, weighing each CCF by the flux of its corresponding exposure. Next, we produced measurements of the signal-to-noise ratio (S/N) by dividing each point by the standard deviation of values located at least 50~\kms\ away from the expected position of the planetary signal.

\subsection{Determination of the systemic velocity of MASCARA-2 and KELT-9}\label{sec:vsys_stellar}
In order to study possible Doppler shifts in the atmospheric signal, we obtained an independent measurement of $v_{\textup{sys}}$ from the stellar spectra. We note that the FIES instrument pipeline does not deliver radial velocity measurements. We therefore used our own analysis, which we briefly describe in this section. We modelled and fitted the radial velocity of the host star in each exposure, using continuum-normalised PHOENIX templates \citep{husser2013}, broadened according to the projected stellar rotational velocity of the star. We Doppler shifted and scaled the template to match each observed spectral order. We combined the radial velocity measurements from each order into an averaged radial velocity measurement, weighting each value by its error. Again, we apply a barycentric correction to the radial velocity of each exposure. Finally, we estimate $v_\textup{sys}$ as the mean and standard deviation of the out-of-transit measurements that were acquired at airmass$<2.0$. From this analysis, we find a value of $v_{\rm sys}=-23.2 \pm 0.4~\kms$ for MASCARA-2, and $v_{\rm sys}=-17.9 \pm 0.4~\kms$ for KELT-9.

\section{Results}
\begin{figure}
\centering
\includegraphics[width=\linewidth]{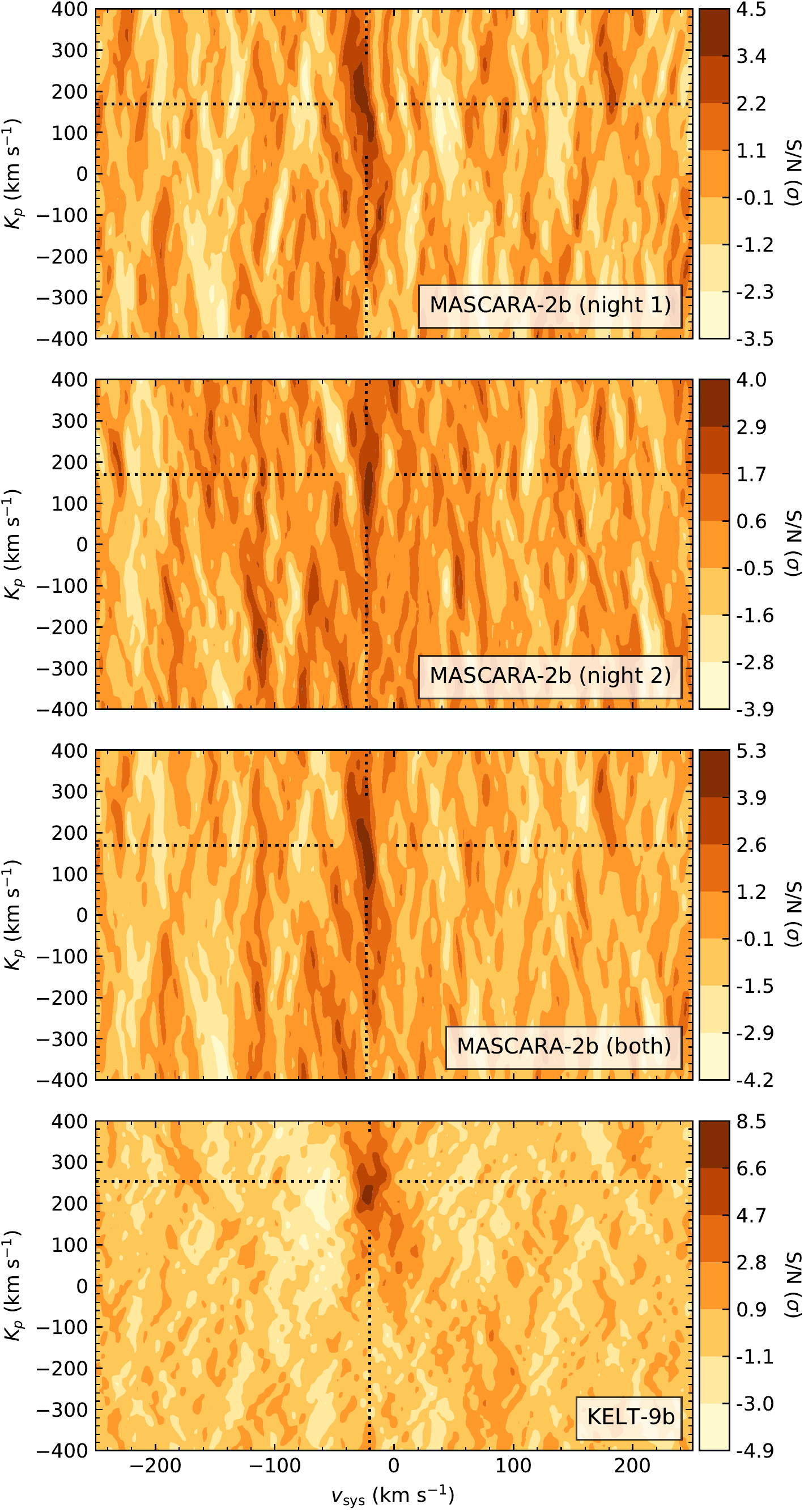}
  \caption{$K_{p}-v_{\rm sys}$ diagrams of Fe II for the transits of MASCARA-2\,b and KELT-9\,b, and for the combination of the two transits of MASCARA-2\,b. The dotted lines indicate the planet radial velocity semi-amplitude and systemic velocity \citep{gaudi2017,lund2017}.}
  \label{fig:kpvsys_all}
\end{figure}

We detect atmospheric Fe II in MASCARA-2\,b in each of the two individual transits, and in KELT-9\,b in the single transit we observed. This is evidenced by the $K_{p}-v_{\rm sys}$ diagrams, shown in Fig.~\ref{fig:kpvsys_all}.
We also include in Fig.~\ref{fig:kpvsys_all} the $K_{p}-v_{\rm sys}$ diagram produced by combining the two transits of MASCARA-2\,b. Table~\ref{tab:peaks} indicates the S/N and location of the Fe II peak in each of the $K_{p}-v_{\rm sys}$ diagrams.
\begin{table}
\centering
{\renewcommand{\arraystretch}{1.4}
\begin{tabular}{cccc}
\hline
\hline
Planet & Night & S/N ($\sigma$) &  \{$K_p,v_{\rm sys}$\} (\kms)  \\
\hline
MASCARA-2\,b & 1 & 4.5 & \{$116^{+146}_{-38},-20^{+2}_{-14}$\} \\
MASCARA-2\,b & 2 & 4.0 & \{$152^{+44}_{-78},-22^{+2}_{-2}$\} \\
MASCARA-2\,b & all & 5.3 & \{$120^{+78}_{-46},-20^{+2}_{-6}$\} \\
\hline
KELT-9\,b & 1 & 8.5 & \{$220^{+16}_{-16},-24^{+2}_{-2}$\} \\
\hline
\end{tabular}}
\caption{Results from the analysis of the Fe II cross-correlation signal, indicating the S/N and velocity pair \{$K_p,v_{\rm sys}$\} with which the signal peaks each night. From the stellar spectroscopy analysis described in Sect.~\ref{sec:vsys_stellar}, we find a systemic velocity of $v_{\rm sys}=-23.2 \pm 0.4~\kms$ for MASCARA-2, and $v_{\rm sys}=-17.9 \pm 0.4~\kms$ for KELT-9.}\label{tab:peaks}
\end{table}

\begin{figure}
\centering
\includegraphics[width=\linewidth]{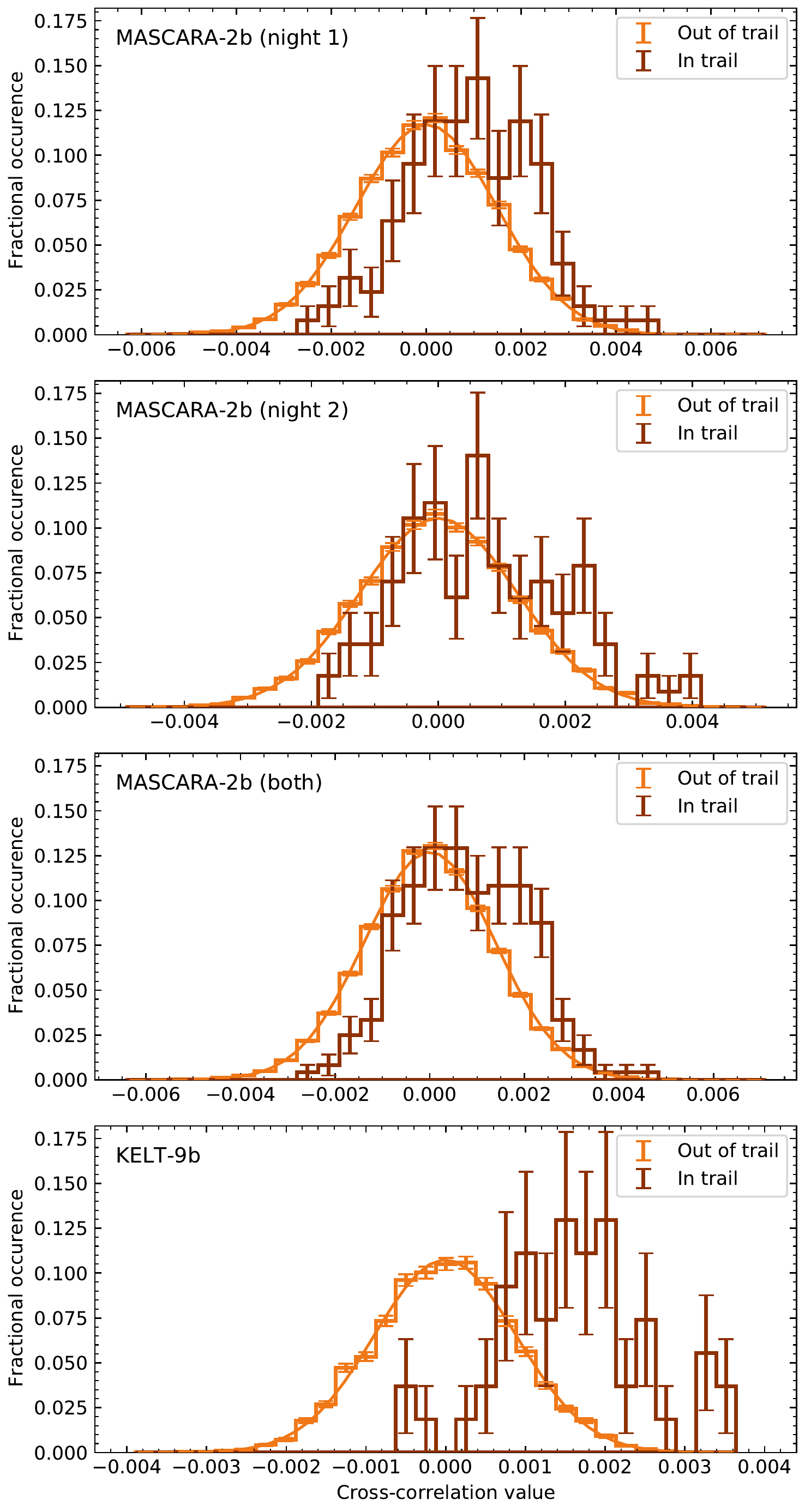}
  \caption{In-trail and out-of-trail Fe II cross-correlation values in the different data sets. We calculated the error bars as the square root of the total number of counts in each bin. The out-of-trail distributions are well approximated by Gaussian functions.}
  \label{fig:welch_tests}
\end{figure}

\subsection{Detection of Fe II in MASCARA-2\,b}
The Fe II signal during the transit of MASCARA-2\,b in Night 1 peaks at $\{K_p,v_{\rm sys}\}=\{116^{+146}_{-38}, -20^{+2}_{-14}\}~\kms$, with $\rm{S/N}=4.5\sigma$, while in Night 2 it peaks at $\{K_p,v_{\rm sys}\}= \{152^{+44}_{-78}, -22^{+2}_{-2}\}~\kms$, with $\rm{S/N}=4.0\sigma$. The quoted uncertainties in the velocities correspond to the $1\sigma$ contours around the peak of the signal. The combination of both nights yields a 5.3$\sigma$ detection, at $\{K_p,v_{\rm sys}\}= \{120^{+78}_{-46}, -20^{+2}_{-6}\}~\kms$. We note that the exact significance of the detection depends on the model template that is used and could, in principle, be optimised by changing parameters such as the model temperature.

These systemic velocities are consistent with those reported by \citet{lund2017} ($v_{\rm sys}=-23.3 \pm 0.3~\kms$), \citet{talens2018} ($v_{\rm sys}=-21.3 \pm 0.4~\kms$), \citet{nugroho2020} ($v_{\rm sys}=-22.06 \pm 0.35~\kms$), and \citet{rainer2021} ($v_{\rm sys}=-24.48 \pm 0.04~\kms$), as well as with the value we find from analysis of the stellar spectra in Sect.~\ref{sec:vsys_stellar}, $v_{\rm sys}=-23.2 \pm 0.4~\kms$, indicating that we do not detect a significant Doppler shift of the Fe II signal from the planet rest frame. While \citet{nugroho2020} do not find a significant Doppler shift of the Fe II signal in MASCARA-2\,b, \citet{casasayasbarris2019}, \citet{stangret2020} and \citet{hoeijmakers2020expres} measure a blueshift of
$\sim 3~\kms$, a value similar to the size of the uncertainties in our measurement.

The orbital parameters of MASCARA-2\,b found in the literature correspond to a radial velocity semi-amplitude of $K_p = 169.3\pm6.2~\kms$ \citep{lund2017} and $K_p = 178\pm19~\kms$ \citep{talens2018}, both also consistent with our results to within the uncertainties. Also consistent with our results are the $K_p$ values obtained directly from the Fe II signal in \citet{nugroho2017} ($K_p = 139.2\pm12.5~\kms$ and $K_p = 165.0\pm3.5~\kms$) and \citet{stangret2020} ($K_p = 155
^{+18}_{-19}~\kms$).

We performed a Welch's $t$-test to reinforce the significance of these detections \citep[e.g.][]{brogi2013,birkby2013,nugroho2017,cabot2019}. We shifted the CCFs to the exoplanet rest frame using the velocities in Table~\ref{tab:peaks}, and we split the in-transit cross-correlation values into two populations: an \textit{in-trail} population and an \textit{out-of-trail} population. The in-trail population consists of the values in the three pixels in each CCF that lie closest to the exoplanet velocity, and the out-of-trail population accommodates all the remaining pixels. Figure~\ref{fig:welch_tests} compares, in histograms, the distribution of the values in these two populations for the different transit observations of MASCARA-2\,b. While the out-of-trail distributions are centered around zero, the in-trail distributions show a significant offset towards positive values. The Welch's $t$-test rules out the hypothesis that the two populations were drawn from the same parent distribution at the 6.9$\sigma$, 5.5$\sigma$ and 8.3$\sigma$ levels for Night 1, Night 2 and the combination of both nights, respectively. We choose to report as the detection significance of our results the more conservative values derived from the S/N measurements in Figure~\ref{fig:ccfs_feii}. This S/N metric relies on the Gaussianity of the cross-correlation values, an assumption supported by the distribution of the out-of-trail values shown in Fig.~\ref{fig:welch_tests}.


\subsection{Detection of Fe II in KELT-9\,b}
The detection of Fe II in the atmosphere of KELT-9\,b peaks at $\{K_p,v_{\rm sys}\}= \{220^{+16}_{-16}, -24^{+2}_{-2}\}~\kms$, with $\rm{S/N}=8.5\sigma$. This $K_p$ value is consistent to the $1\sigma$ level with the value that \citet{hoeijmakers2019} find from an analysis of the Fe II signal. \citet{yanhenning2018} derive a larger value using the $H_\alpha$ signal, $K_p = 268.7^{+6.2}_{-7.3}~\kms$. We note that the derived $K_p$ is a combination of the orbital $K_p$ ($K_p = 253.9\pm7.3~\kms$, \citealt{gaudi2017}) and complex atmospheric dynamics. As a consequence, we expect that the $K_p$ measurements from different species might be dissimilar \citep[e.g.][]{wardenier2021,rainer2021,prinoth2022,kesseli2022}.

Due to the high projected stellar rotational velocity of KELT-9 ($v\sin{i_\star} \simeq 110~\kms$), measurements of its systemic velocity are challenging \citep{gaudi2017}. Values in the literature based on stellar spectroscopy show significant discrepancies: $v_{\rm sys} = -20.567\pm0.1~\kms$ \citep{gaudi2017}; $v_{\rm sys} = -17.74 \pm 0.11~\kms$ \citep{hoeijmakers2019}; $v_{\rm sys} = -19.819\pm0.024~\kms$ \citep{borsa2019}; and $v_{\rm sys}=-17.86 \pm 0.044~\kms$ \citep{paiasnodkar2022}\footnote{This value corresponds to the most precise measurement in \citet{paiasnodkar2022}, derived from PEPSI spectra. The authors also report measurements using other spectrographs: $v_{\rm sys}=-17.15 \pm 0.11~\kms$ from HARPS-N, and $v_{\rm sys}=-18.97 \pm 0.12$ from TRES.}. From stellar spectroscopy, we find a value of $v_{\rm sys}=-17.9 \pm 0.4~\kms$. In our analysis, the systemic velocity of the Fe II signal appears blue-shifted by a few \kms\ with respect to these values, which might be due to high-velocity (3--$6~\kms$) winds flowing across the terminator of KELT-9\,b. However, we caution that the uncertainties in our measurement of $v_{\rm sys}$ are large.

As in the case of MASCARA-2\,b, we compared the in-trail and out-of-trail populations. Figure~\ref{fig:welch_tests} clearly highlights that the distribution of the in-trail values is notably shifted with respect to the out-of-trail population. In this case, a Welch's $t$-test rejects the null hypothesis at the $8.6\sigma$ confidence level.

\subsection{The search for Fe I in MASCARA-2\,b and KELT-9\,b}
\begin{figure}
\centering
\includegraphics[width=\linewidth]{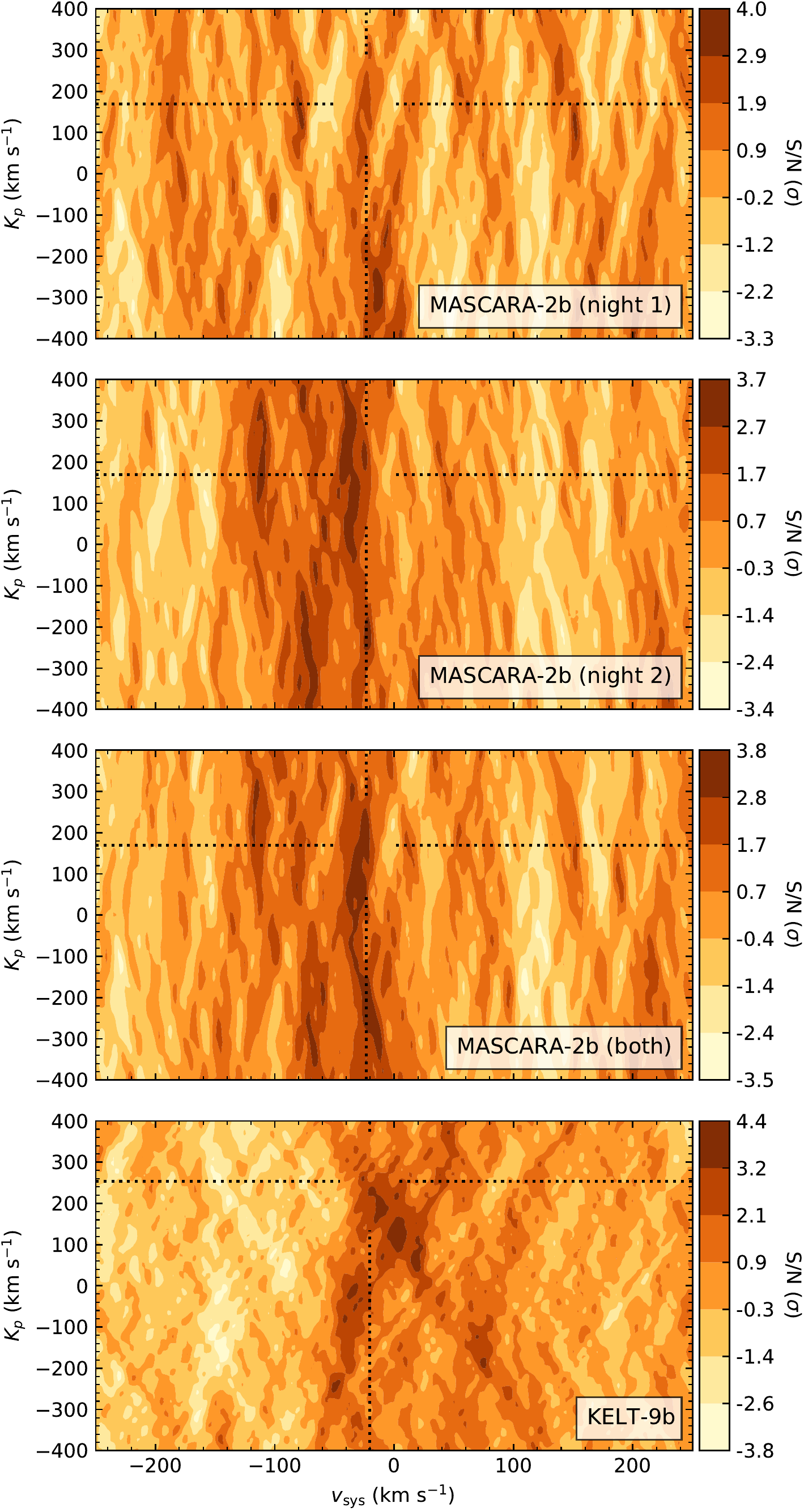}
  \caption{Same as Fig.~\ref{fig:kpvsys_all}, but for Fe I.}
  \label{fig:kpvsys_fei}
\end{figure}

The $K_{p}-v_{\rm sys}$ diagrams of Fe I, in Fig.~\ref{fig:kpvsys_fei}, reveal a signal at the $\sim 3 \sigma$ level in the KELT-9\,b and the combined MASCARA-2\,b data consistent with the respective velocities of the planets. Although it is plausible that these signals are of planetary origin, we argue that we need more data before claiming a detection of Fe I due to the presence of additional high-S/N signals in the $K_{p}-v_{\rm sys}$ diagrams. In particular, the Fe I $K_{p}-v_{\rm sys}$ diagram from combining both nights of MASCARA-2\,b data (third panel in Fig.~\ref{fig:kpvsys_fei}) presents a second strong peak at a $K_p$ opposite of that of the planet. The Fe I $K_{p}-v_{\rm sys}$ diagram of KELT-9\,b exhibits a multiple-peak structure that starts at a velocity pair consistent with that of the planet, and extends towards significantly higher $v_{\rm sys}$ and lower $K_{p}$ values.

In high resolution spectroscopy, and under the assumption of photon-dominated noise, the S/N scales with the amplitude of the transmission signal and the square root of the number of resolved lines \citep{snellen2015,birkby2018}.
Fe I presents many more absorption lines than Fe II at the wavelengths covered by FIES. However, the average line depth of Fe I in MASCARA-2\,b and KELT-9\,b is an order of magnitude smaller than that of Fe II \citep{hoeijmakers2019,hoeijmakers2020expres}, contrary to predictions from isothermal models that assume hydrostatic and chemical equilibrium. This factor is key in making Fe I more challenging to detect than Fe II in our data.

\section{Discussion and Conclusions}
In this work, we analyse high-resolution spectroscopic data of two transits of MASCARA-2\,b and a transit of KELT-9\,b, observed with FIES at the Nordic Optical Telescope. We find strong absorption by atmospheric Fe II in each of the two transits of MASCARA-2\,b (S/N of 4.5$\sigma$ and 4.0$\sigma$), and in the transit of KELT-9\,b ($8.5\sigma$).

Our results demonstrate the feasibility of studying atmospheres at high spectral resolution with FIES/NOT. FIES has the stability and precision required for characterisation of exoplanet atmospheres, and it benefits from a broad spectral range that reaches wavelengths as long as 7000--8820~\AA, which many of the other optical spectrographs that are often used in similar studies cannot do. Such broad spectral range is particularly relevant for high-resolution cross-correlation spectroscopy, where the achieved S/N scales, to first order, with the square root of the number of planet lines in the spectrum \citep{snellen2015}.

Additionally, we note that the data analysed in this work is prior to the recent recoating of the FIES collimator and folding mirrors. With the new mirror coatings, FIES has become 60--140\% more efficient (J. Telting, priv. comm.).

More generally, this proof-of-concept study opens the door for characterisation of exoplanet atmospheres with other modest-size telescopes equipped with similar spectrographs. There is a large suite of high-resolution spectrographs on 2-m class telescopes whose potential for characterisation of exoplanet atmospheres remains largely unexplored, such as FOCES on the 2-m Fraunhofer Telescope at Wendelstein Observatory \citep{pfeiffer1998}. As NASA's \textit{TESS} mission keeps delivering numerous targets transiting bright stars, the availability of multiple spectrographs capable of this work will be crucial to exploring the composition and dynamics of exoplanet atmospheres and understanding the systematics of different instruments.

\begin{acknowledgements}
We thank the anonymous referee for a prompt and detailed report that improved the quality of our work. We also thank the staff at the Nordic Optical Telescope, who made the observations presented in this work possible. This study is based on observations made with the Nordic Optical Telescope, owned in collaboration by the University of Turku and Aarhus University, and operated jointly by Aarhus University, the University of Turku and the University of Oslo, representing Denmark, Finland, and Norway, the University of Iceland and Stockholm University at the Observatorio del Roque de los Muchachos, La Palma, Spain, of the Instituto de Astrofisica de Canarias. A.B.-A. gratefully acknowledges support from "la Caixa" Foundation (ID 100010434), under agreement LCF/BQ/EU19/11710067. A.W.M. is supported by the NSF Graduate Research Fellowship grant no. DGE 1752814.
\end{acknowledgements}


%
%
\bibliography{sample631}{}
\bibliographystyle{aa}

\end{document}